\documentclass[fleqn,10pt,twocolumn]{wlscirep}
\usepackage{setspace}
\usepackage[squaren]{SIunits}

\title{Beyond conventional photon-number detection with click detectors}

\author[1,*]{Johannes Kr\"oger}
\author[1,+]{Thomas Ahrens}
\author[2]{Jan Sperling}
\author[2]{Werner Vogel}
\author[1]{Heinrich Stolz}
\author[3,+]{Boris Hage}
\affil[1]{University of Rostock, Semiconductor Optics Group, Physics Institute, Rostock, 18059, Germany}
\affil[2]{University of Rostock, Theoretical Quantum Optics Group, Physics Institute, Rostock, 18059, Germany}
\affil[3]{University of Rostock, Experimental Optics Group, Physics Institute, Rostock, 18059, Germany}

\affil[*]{johannes.kroeger@uni-rostock.de}

\affil[+]{these authors contributed equally to this work}

\begin{abstract}
Photon-number measurements are a fundamental technique for the discrimination and characterization of quantum states of light. Beyond the abilities of state-of-the-art devices, we present measurements with an array of 100 avalanche photodiodes exposed to photon-numbers ranging from well below to significantly above one photon per diode. Despite each single diode only discriminating between zero and non-zero photon-numbers we are able to extract characteristic information about the quantum state. We demonstrate a vast enhancement of the applicable intensity range by two orders of magnitude relative to the standard application of such devices. It turns out that the probabilistic mapping of arbitrary photon-numbers on a finite number of registered clicks is not per se a disadvantage compared with true photon counters. Such detector arrays can bridge the gap between single-photon and linear detection, by directly using the recorded data, without the need of elaborate data reconstruction methods.
\end{abstract}
\begin{document}
\doublespacing
\flushbottom
\maketitle
\section*{Introduction}

Quantum optics, quantum communication, and quantum computation depend on the ability to discriminate quantum states of a system, based on measurements.
For light fields several proposals have been made to obtain statistical information of the quantum state and some of these proposals, e.g. reconstructing quasiprobabilities and photon number distributions, became standard tools for studying quantum properties of light~\cite{loudon2000quantum,vogel2006quantum,agarwal2012quantum,vogelrisken89}.
In the last years remarkable progress in experiments with microscopic and mesoscopic intensities has been made. In parallel, increasing demands on photon detectors during this progress caused more sophisticated photon counting techniques to be developed~\cite{josab2014,migdall2013single,Hadfield2009}.
Very promising examples are superconducting transition-edge and nano\-wire detectors, visible light photon counters, frequency up-conversion and avalanche photodiodes~\cite{Irwin95,Waks2004,stahl2005cryogenic}.

Some of these detectors, e.g. an avalanche photodiode in Geiger mode and a single superconducting nanowire,
own a single photon sensitivity but provide only a binary `click' for any number of absorbed photons. Hence, these devices are only capable of discriminating between zero and one or more photons. 
The only way of retrieving photon number information is by either distributing the signal onto multiple click detectors at the same time (arrays of avalanche photodiodes, nanowire detectors generally have a multitude of wires)~\cite{Jiang2007} or by the piecewise detection of the splitted and delayed signal pulse by one or few click detectors (time multiplexing)~\cite{haderka2004,Fitch2003,Bartley2012}.


This class of so called pixelated photon detectors (PPDs), needs a small quotient between number of incident photons and detector number/time bins, as to avoid multiple photons on a single detector/time bin (henceforth referred to as pixels). It has been shown, that the number of photons has to be roughly two orders of magnitude smaller than the total number of pixels on the PPD, so that the recorded click numbers can be approximated to be photon numbers\cite{sperling12a}.
Thus, even the most sophisticated experimental setup 
allowed proper characterization of a light field only in the few-photons-detection regime~\cite{Jiang2007}. 
Regardless of that, PPDs used within the small-photon-number-per-pixel condition~\cite{sperling12a} are utterly capable of performing true photon counting detector tasks, such as single shot photon counting and quantum state measurements~\cite{Afek2009}.

Based on the recently developed theoretical click counting distribution~\cite{sperling12a,sperling12b,sperling15}, the exact click statistics can be calculated, for any quantum state at any intensity, allowing to infer the state of a light field in a similar way as from the usual photon statistics.
\begin{figure}[t]
\centering
\includegraphics[width=\columnwidth]{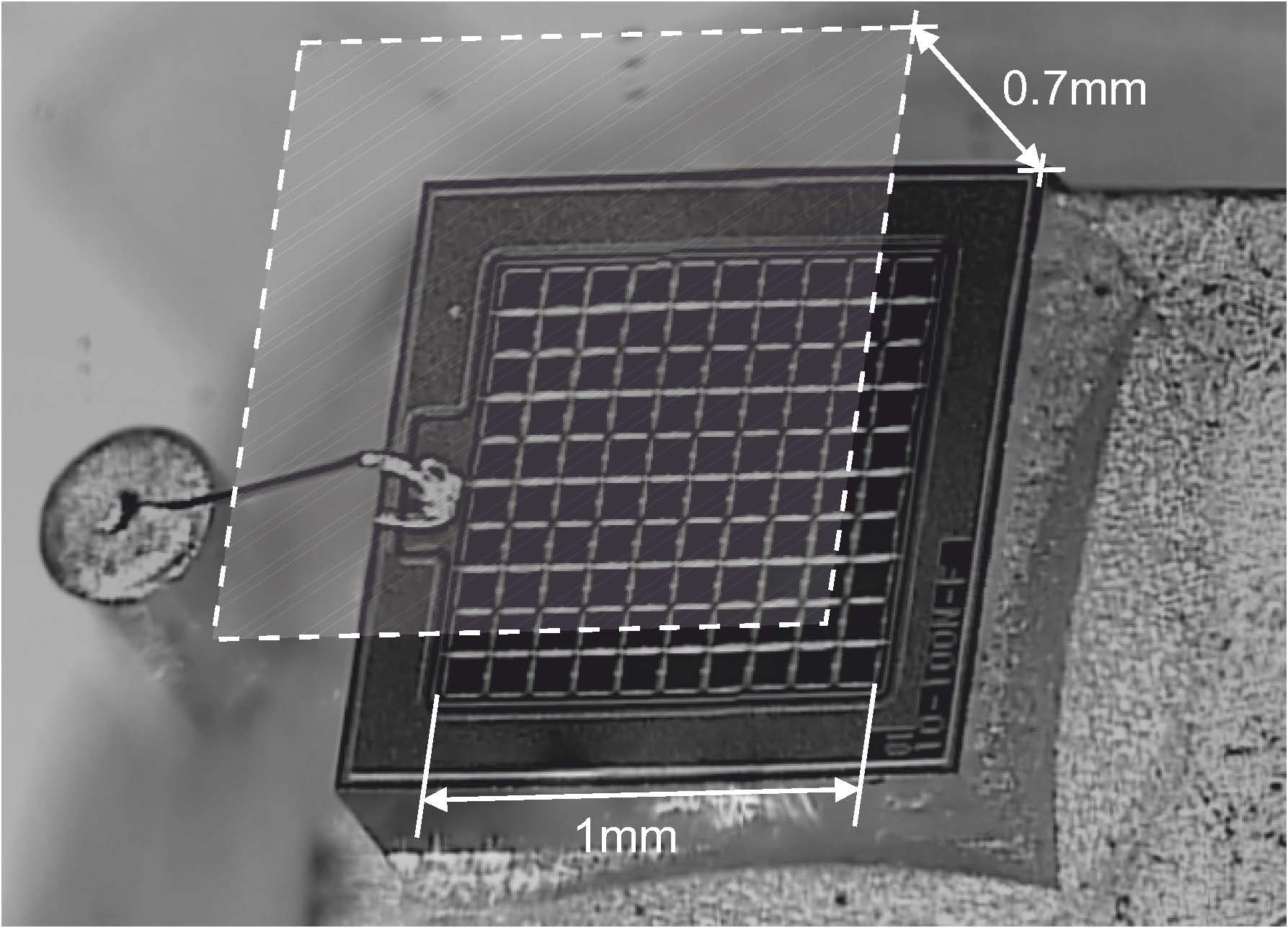}
\caption{{\bfseries Click detector array.} A 10 $\times$ 10 array of single photon sensitive click detectors connected in parallel, which constitute the MPPC diode produced by Hamamatsu. Cap window indicated.}
\label{pic:foto}
\end{figure}
In this paper we demonstrate measurements with a PPD consisting of 100 single avalanche photodiodes in Geiger mode, where single detectors elements are triggered by photon numbers ranging from well below to explicitly above one photon per pixel, in average. Even though information loss occurs when multiple photons trigger the same pixel, we show that significant statistical information about the quantum state is still preserved within the click statistics. Although we confine our investigations to an array of avalanche photo diodes, our results and conclusions related to the click counting distribution in general should apply to any other system of click detectors, whereas our approach on detector imperfections, such as crosstalk and dark counts, may only be valid for avalanche-photodiode-based PPDs.

\section*{Methods}
For a successful observation of click statistics at high intensities a number of essential requirements have to be fulfilled. 
At first we will outline the theoretical framework of this paper, followed by the requirements to our detector and PPDs in general and the experiment setup. Finally, we explain how we obtained the click statistics from the detector output.

\subsection*{Click counting distribution}
The general click counting distribution for a PPD with $N$ homogeneously illuminated click detectors and any quantum state at any intensity is given by\cite{sperling12a}:
\begin{align}\label{eq:binomial-click}
c_k&=\left<:\right.\binom{N}{k}\left(\mathrm{e}^{-\frac{\eta\hat{n}+\nu}{N}}\right)^{N-k}\left(1-\mathrm{e}^{-\frac{\eta\hat{n}+\nu}{N}}\right)^k\left.:\right>\,,
\end{align}
where $c_k$ is the probability to detect $k$ clicks, $\hat n$ is the photon number operator, $\eta$ is the detection efficiency/loss and $\nu$ is the dark count rate, which is assumed to be Poisson-distributed.
For this work, the click distribution for coherent states is the most relevant. In that case eq.~\ref{eq:binomial-click} simplifies to a binomial distribution,
\begin{align}\label{eq:coherent-click}
	c_{k}(\alpha)&=\binom{N}{k}\left(\mathrm{e}^{-|\alpha|^2/N}\right)^{N-k}\left(1-\mathrm{e}^{-|\alpha|^2/N}\right)^k\,,
\end{align}
where $1-\mathrm{e}^{-|\alpha|^2/N}$ is the probability for a click at a single pixel and $\alpha$ is the coherent amplitude.

Note that, the click counting distribution and its consequent inferences are based on the assumption that all pixels are only triggered once during one measurement period. Hence the light source is restricted to be pulsed, with pulsewidths smaller and repetition times bigger than the dead time of the pixels. To our knowledge, a click counting distribution for continuous light has not been formulated by now. The application of PPDs to continuous light, however, is practiced within the small-photon-number-per-pixel condition, where the PPD is used as a true photon counter~\cite{Renker2009207,Du2008396}.

\subsection*{Crosstalk}
Additional classical noises affect the click statistics by a convolution of noise statistics with the statistics of pixels triggered by light,
\begin{align}
	c_k&=\sum_{i=0}^kc^\mathrm{light}_{i}c^\mathrm{\vphantom{light}noise}_{k-i}\,.
\end{align}
In our experiment crosstalk~\cite{1503521,4447292} had the greatest impact to the click statistics. To describe crosstalk our model is based on two simple assumptions: First, crosstalk is homogeneously distributed over the remaining active pixels, meaning that every pixel which has not been triggered has an equal probability to be crosstalk triggered by a previously clicked pixel. Second, crosstalk triggered pixels can trigger additional crosstalk pixels. Following these assumptions we only need to introduce one crosstalk parameter, e.g. the probability $\chi$ of a certain pixel to be triggered by one previously clicked pixel, and the probability to trigger $k$ clicks by crosstalk of $N_\mathrm{P}$ previously clicked pixels and $N_\mathrm{A}$ available pixels is binomially distributed in $k$ (see supplement):
\begin{align}\notag
	C_{k}^\mathrm{cross}(N_\mathrm{A},N_\mathrm{P},\chi)&=\binom{N_\mathrm{A}}{k}(1-(1-\chi)^{N_\mathrm{P}})^k\\
	&\hphantom{==}\times(1-\chi)^{N_\mathrm{P}(N_\mathrm{A}-k)}\,.
\end{align}
Allowing crosstalk clicked pixels to trigger additional clicks via crosstalk leads to the probability of $k$ crosstalk clicks being triggered by $a_0$ initial light clicks:
\begin{align}\label{eq:crosstalk}
	c_{k}^\mathrm{cross}(a_0,\chi)=\sum_{\mathbf a=\mathrm{IP}(k)}\prod_{a_i}C^\mathrm{cross}_{a_i}(N_\mathrm{A}-\textstyle\sum_{l=0}^{i-1}a_l,a_{i-1},\chi)\,,
\end{align}
where $\mathbf a=(a_1,\dots,a_i,\dots,a_n)=\mathrm{IP}(k)$ are the integer partitions of $k$ and $a_0$ is the number of initial light clicks.
The probability to detect overall $m$ clicks including coherent statistics~\eqref{eq:coherent-click} and crosstalk~\eqref{eq:crosstalk} is then:
\begin{align}
c_m(\alpha,\chi)=\sum_{k=1}^mc_k(\alpha)c_{m-k}^\mathrm{cross}(k,\chi)
\end{align}

\subsection*{Nonclassicality indication}
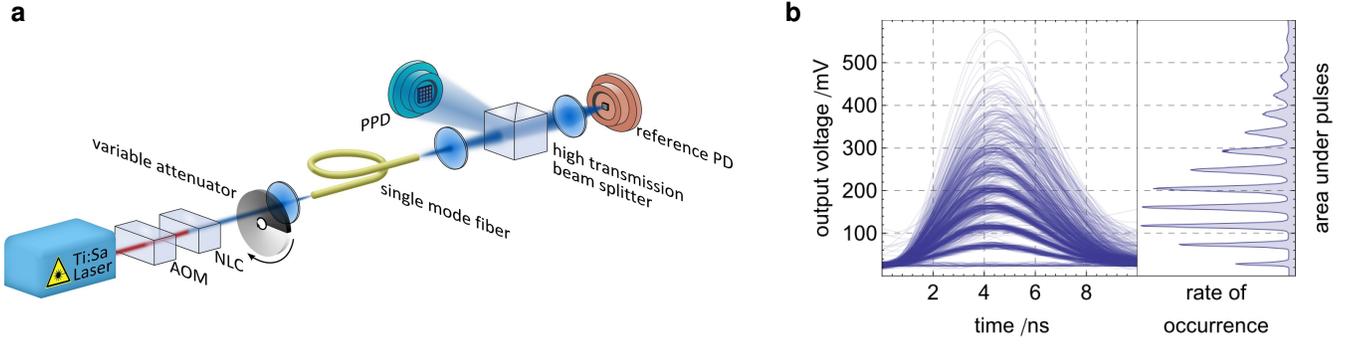
\begin{figure*}[t]
\parbox{0.6\textwidth}{
\includegraphics[width=.55\textwidth]{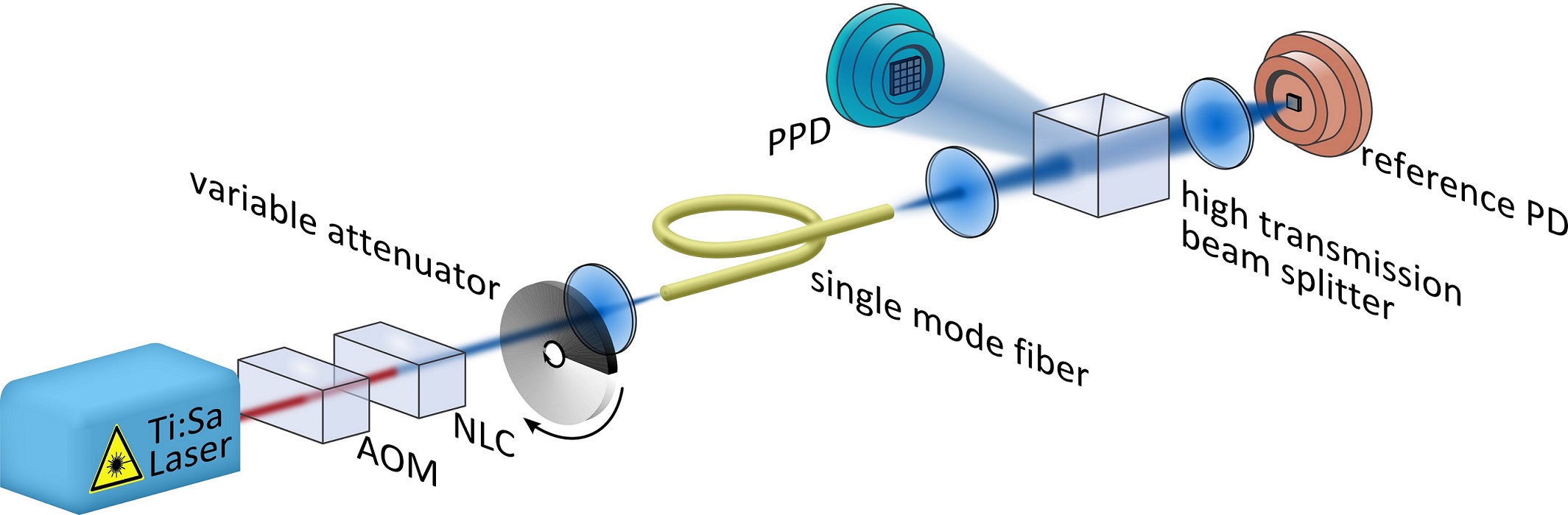}
}
\parbox{0.4\textwidth}{
\includegraphics[width=0.4\textwidth]{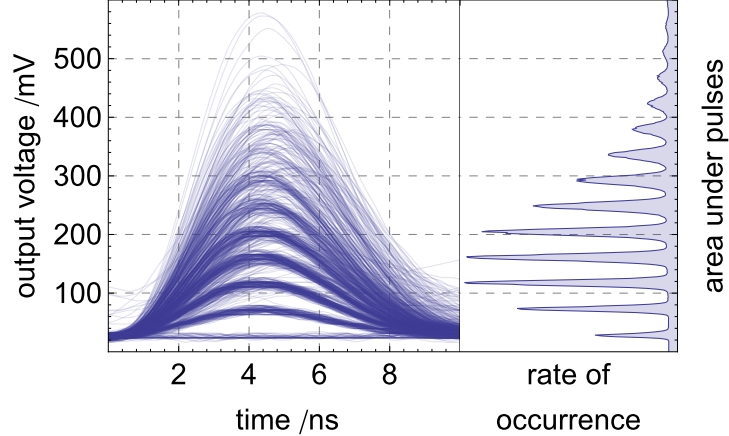}
}
\begin{picture}(0,0)
\setlength{\unitlength}{.1\textwidth}
\put(0.05,2.4){\makebox(0,0)[bl]{\bfseries\textsf{a}}}
\put(5.9,2.4){\makebox(0,0)[bl]{\bfseries\textsf{b}}}
\end{picture}
\caption{{\bfseries Applied click detection.} {\bfseries\textsf a}, Setup: The acousto-optic modulator (AOM) decreases the repetition rate of the Ti:Sa laser and a nonlinear crystal (NLC) halves the wavelength to 400\,nm for efficient detection at the pixelated photon detector (PPD). {\bfseries\textsf b}, Data sampling: The area under each pulse of the output signal is proportional to the number of clicking pixels within one light pulse. The occurrence histogram of the pulse areas contains all information about the click statistics.
}
\label{pic:Setup}%
\end{figure*}
Indicators which are used reveal distinct quantum properties in photon or click statistics often quantify coherent states as bordering between classical and quantum states.
A common nonclassicality indicator for photon statistics is the Mandel-Q parameter~\cite{Mandel:79}, $Q_\mathrm{M}=\langle(\Delta n)^2\rangle/\left<n\right>-1$, with $\langle(\Delta n)^2\rangle$ and $\left<n\right>$ being variance and mean of the photon statistics. $Q_\mathrm{M}$ is positive semidefinite for classical states, zero for states with Poisson-distributed photon statistics, such as coherent states, and can only be below zero for quantum states with subpoisson photon statistics, which have no classical analogon.
For click statistics of PPDs a similar parameter has been suggested, namely the binomial Q parameter\cite{sperling12b}:
\begin{align}\label{eq:qbinomial}
Q_\mathrm{B}&=\frac{\langle(\Delta c)^2\rangle}{\left<c\right>(1-\frac{\left<c\right>}{N})}-1\,,
\end{align}
with the mean value $\left<c\right>$ and the variance $\langle(\Delta c)^2\rangle$. Similar to $Q_\mathrm{M}$, $Q_\mathrm{B}$ is positive semidefinite for classical fields, zero for coherent fields and can only be below zero for quantum states.
Note that, the application of $Q_\mathrm{M}$ to click statistics can be justified within the small-photon-number-per-pixel condition, but it technically indicates nonclassicality for coherent states at any intensity, with the magnitude $\mathrm{e}^{-|\alpha|^2/N}-1$.




\subsection*{Detector performance and experiment setup}
\begin{figure*}[t]
\includegraphics[width=\textwidth]{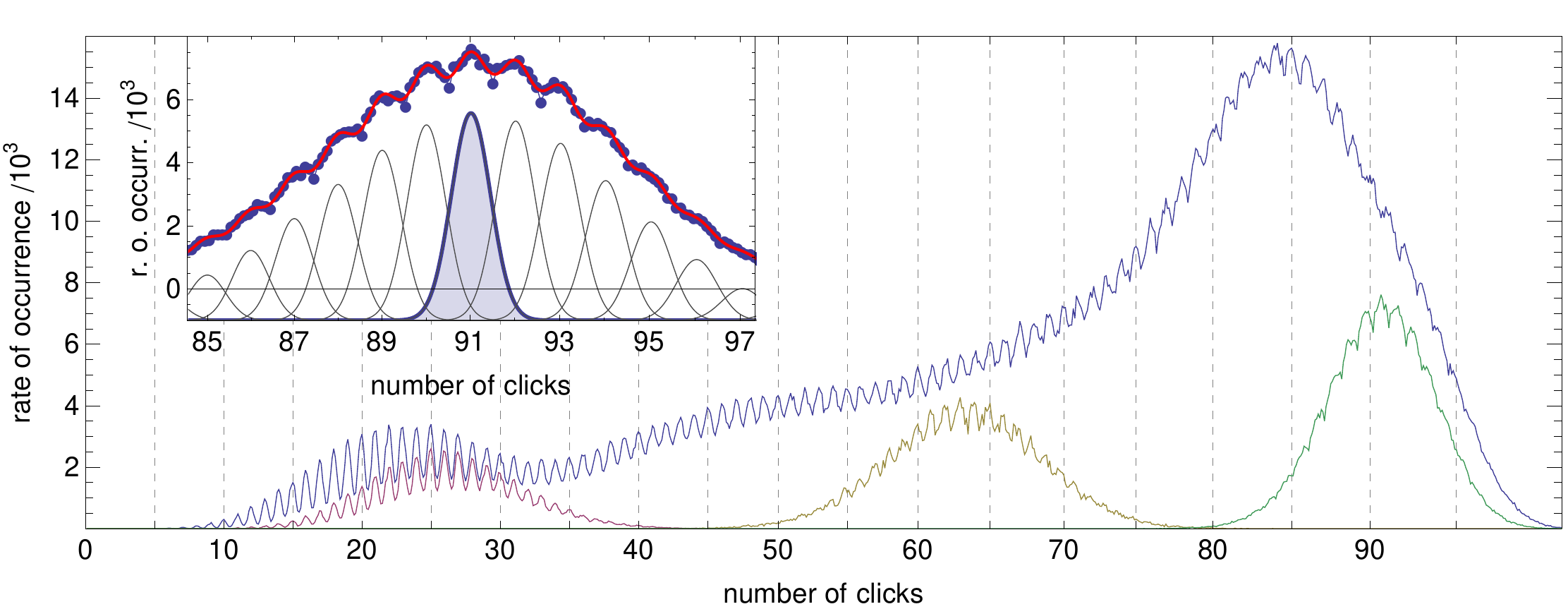}
\caption{{\bfseries Click number resolution.} The blue curve is a compilation of detector output raw data at various intensities, whereas the red, yellow and green curves are raw data at distinct and narrow intensities. The abscissal coordinate of a particular click number can be assigned by simply counting the peaks in the blue curve.
This is essential for the assignment of click numbers to occurrence histograms with narrow intensity ranges. Inset: Decomposition of the histogram plot into the Gaussian contributions of the different click numbers.
}
\label{pic:AlleKlicks}%
\end{figure*}
For the investigation of a light field in a single spatial mode it is of no interest how much a particular pixel contributed to the click statistics, hence it is not necessary to process the output of every pixel individually. 
The PPD system we used consists of a MPPC-Diode with 100 single avalanche photodiodes in Geiger mode, connected in parallel, designed and produced by Hamamatsu Photonics~\cite{yamamoto2007} combined with a unique low-noise transimpedance amplifier carefully designed in house. 
Together with good temperature and reverse voltage stability ($\approx0.1\,\kelvin$ and $\approx1\milli\volt$) we were able to resolve almost all of the 100 possible clicks (see figure~\ref{pic:AlleKlicks}).



The light source in our experiment was a Ti:Sa-fs-laser with a center wavelength of $\lambda=800\,\nano\meter$.
Since the pulse repetition time of the Ti:Sa-laser ($\Delta t_\mathrm{Ti:Sa}\approx12.5\,\nano\second$) is smaller than the dead time of the PPD pixels ($t>80\,\nano\second$), triggered pixels could not recover between two fs-pulses.
Therefore, an acousto-optic modulator (AOM) with a sharp transition was used to release pulses with a larger repetition time ($\Delta t_\mathrm{AOM}\approx2\,\micro\second$) into the second harmonic generating nonlinear crystal NLC ($\lambda_\mathrm{SHG}=400\,\nano\meter$), which converted the wavelength of the source light to the high sensitivity range of the PPD.

A variable attenuator served as a manipulator for the source intensity and a single mode fibre reduced the multimode dispersion on the PPD.
The low reflectance of the beam splitter (BS) and the great divergence of the beam provided a poissonian photon statistics at the PPD and furthermore ensured that every pixel is illuminated equally. Due to the focusing lens after the beam splitter, the reference photo diode (reference PD) can be operated linearly.
The reference PD recorded the intensity individually for every pulse which allows discrimination of the pulse intensity at the PPD via post selection.

It is important to mention, that single shot photon counting, meaning the exact determination of discrete photon numbers from light pulse to light pulse, as it is currently practiced with PPDs like ours, is still permitted within the small-photon-number-per-pixel condition. Electronic noise and fluctuations in gain lead to overlapping peaks in the rate of occurrence histogram in figur~\ref{pic:Setup}b for click numbers above a certain value. However, the limit of single shot resolvable clicks is still greater than -- and thus does not interfere with -- the limit in photon numbers, so that two photons never hitting the same pixel can be assumed savely.


\subsection*{Click statistics acquisition}
For a -- conceptionally -- ideal PPD the electric charge which is emitted per clicking pixel would be equal for all pixels, thus in our experiment only depending on the capacitance of the APD pixels and the applied reverse voltage (commonly taken together as "gain").
Therefore, the electric charge of simultaneously clicking pixels should be an integer multiple of the electric charge of a single click (the charge of a single APD avalanche).
As seen in figure~\ref{pic:Setup}{b} the temporal progression of the output of the PPD is a pulse of several ns duration. The area under each output pulse is proportional to the electric charge emitted by the pixels, which are triggered by the corresponding light pulse. Hence, a frequency distribution of the areas under the pulses (AuP distribution, see figures~\ref{pic:Setup}b and~\ref{pic:AlleKlicks}) contains all information about the click statistics.

Due to the -- actually -- random nature of the avalanche multiplication process and the likely varying APD capacities the emitted charges per click are statistically fluctuating and can safely be approximated by a Gaussian distribution. Even though that the tails of neighbouring peaks in the AuP distribution overlap at high click numbers, the click statistics can be easily obtained by fitting the AuP distribution to a sum of Gaussians in the following way (see Inset of figure~\ref{pic:AlleKlicks}):
\begin{align}
	f(x)&=\sum_{i=0}^N\frac{A_i}{\sqrt{\sigma_0^2+i\sigma_1^2}}\exp\left[-\frac{1}{2}\frac{\left(x-i\Delta x\right)^2}{\sigma_0^2+i\sigma_1^2}\right]\,,
\end{align}

\noindent where $x$ is an arbitrary abscissa scale, $\Delta x$ corresponds to the gain and is therefore responsible for the distance between neighbouring peaks, $\sigma_0$ corresponds to electronic noise/width of the zero-click-peak, $\sigma_1$ corresponds to the noise of one click and $A_i$ is the area under the $i$th peak. The AuP peak for exactly $i$ clicks is a multiple convolution of the 0-clicks-peak (electronic noise) and $i$ times the one-click-peak (Gaussian distribution for one click).
Typical values were $\sigma_0/\Delta x=0.18$ and $\sigma_1/\Delta x=0.0037$.
The click statistics are then obtained by $c_k=A_k/\sum_{i=0}^NA_i$.

As stated above, the single-shot click counting resolution is limited by the noise ($\sigma_0$ and $\sigma_1$) and is around $\leq10$ clicks for our detector. The single-shot photon counting ability however is limited  small-photon-number-per-pixel condition (as stated in the introduction), which is less than 5 clicks for our detector. Therefore, the overlapping peaks in an AuP distribution and the above described click statistics acquisition procedure do not interfere with the single-shot photon counting abilities of our detector.



\section*{Results}
\begin{figure}[t]
\centering
\includegraphics[width=\columnwidth]{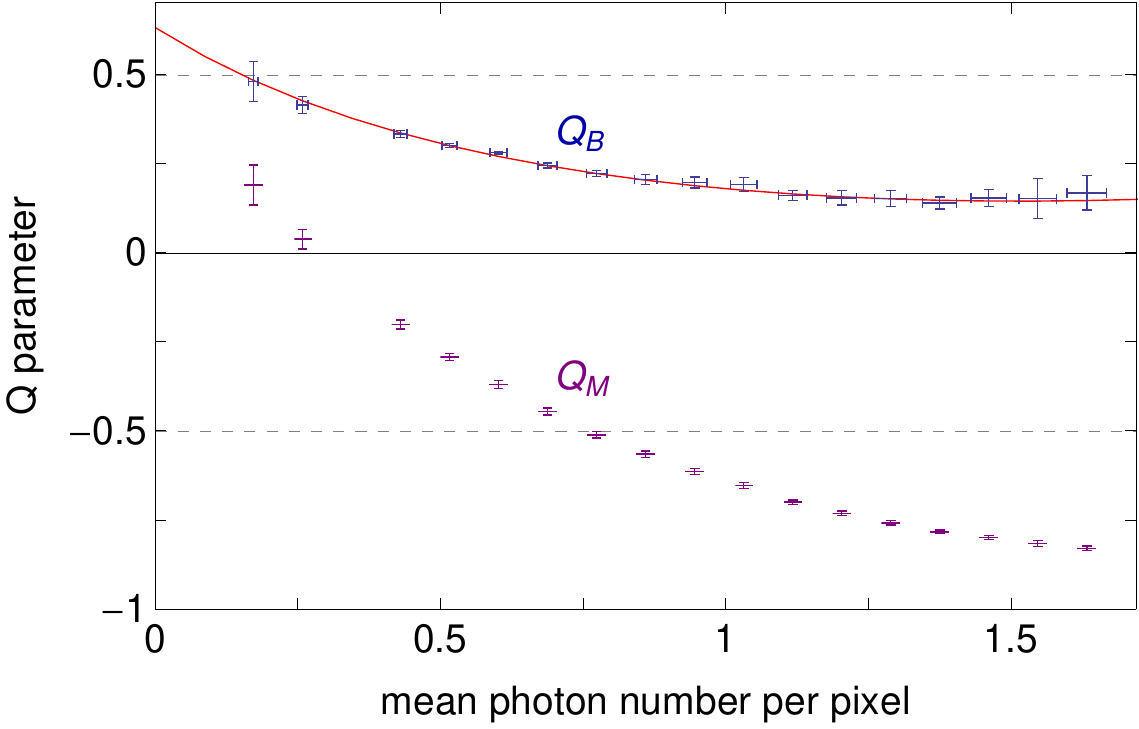}
\caption{{\bfseries Q-parameter.} $Q_\mathrm{B}$ (blue) and $Q_\mathrm{M}$ (violett) and their error extracted from our data, as well as the theoretic behavior of $Q_\mathrm{B}$ (solid red line).
$Q_\mathrm{M}<0$ falsely indicates nonclassicality of a coherent source.
Crosstalk and preclicked pixels affect the click statistics and convolve the binomial statistics with classical noise, thus producing significantly positive values of $Q_\mathrm{B}$. Crosstalk parameter: $\chi=0.25\%$.
}
\label{pic:QParameter}
\end{figure}

\subsection*{General considerations}
The general click counting distribution~\eqref{eq:binomial-click} is capable of quantifying the properties of a any quantum state at any intensity, allowing to infer the state of a light field in a similar way as from the actual photon statistics without the need of reconstructing the photon statistics. With our results we show that the applicable intensity range of click detectors can be increased by nearly two orders of magnitude, overcoming the prevalent small-photon-number-per-pixel condition, with no hard intensity cut-off and only limited by the measurement precision and significance of the observed quantity.

Considering equation \eqref{eq:coherent-click} it is easy to see that the nearly 100 resolved clicks of our detector correspond to more than 3 photons per pixel in average, while still being able to access the click statistics. Unfortunately classical noises limit the significance of our measurements, thus restricting our analysis of the Q-parameters to an average photon number of 1.5 per pixel. Future developments may extend this limitation.

Note that, the inversion of click statistics to photon number statistics, which is occasionally practiced in the low intensity regime~\cite{worsley:09}, technically has to fail to produce the correct results, since photon numbers can not be mapped bijectively to click numbers. At higher intensities this can even result in negative photon number probabilities (see supplement). This especially is the case when photon numbers are involved, which are larger then the number of pixels of the PPD.

\subsection*{Nonclassicality indication}
With the variable attenuator (see fig.~\ref{pic:Setup}a) we were able to alter the signal intensity and recorded a series of click distributions like to those shown in Figure \ref{pic:AlleKlicks}. We applied both, the Mandel Q-parameter $Q_\mathrm{M}$ and the binomial Q-parameter $Q_\mathrm{B}$, to the click statistics and plotted them  vs. the mean number of photons per pixel (fig.~\ref{pic:QParameter}).
As expected, $Q_\mathrm{M}$ applied to click statistics -- without critical appraisal -- falsely indicates subpoisson light with a strong magnitude at high photon numbers per pixel.

In contrast to that, $Q_\mathrm{B}$ is always greater than zero and does not indicate nonclassicality. Admittedly, instead of the $Q_\mathrm{B}$ value of zero, as would be expected for any amplitude of a coherent source, $Q_\mathrm{B}$ shows significantly positive values ranging from 0.5 to 0.1, from low to moderate intensities. The origin of this behaviour is additional classical noise which contributes to the click statistics and which we identified to be crosstalk (see section: Classical noises). Including classical noise to our detector model and fitting the crosstalk parameter and an intensity scaling factor to our data we are able to explain the behavior of $Q_\mathrm{B}$. 
It is important to mention, that although we fitted the crosstalk parameter it is technically a fixed value determined for $\lim_{|\alpha|\rightarrow0}Q_\mathrm B$ (see section: Classical noises). In theory, the intensity scaling factor could as well be determined separately by the reference photodiode in our setup, but due to the many sources of uncertainty (efficiencies, lenses, beam splitter, beam divergence) we were unable to achieve that with sufficient significance. Furthermore the intensity scaling factor has no influence to the deviation of $Q_\mathrm B$ form 0.

Note that, the greatest deviation between including and excluding classical noise in the detector model is at low intensities, where click detector systems are currently used due to the small-photon-number-per-pixel condition.

\subsection*{Classical noises}
The major detector imperfections in our experiment are problems well known for avalanche based photon detectors, namely dark counts, preclicked pixels and crosstalk. Since we used a pulsed light source with pulse widths much smaller than the dead time of the PPD pixels and we were restricting the measurement time to the width of one output pulse of the detector, alterpulsing -- clicks subsequently following another click of the same pixel -- have no effect to our measurements and results. At first we will address effects based on dark counts, which includes effects related to preclicked pixels, before we consider the influence of crosstalk.

The impact of dark counts to the click statistics must be differentiated into two separate effects. Dark counts which occur during the measurement period for one light pulse are taken into account in the click counting statistics and therefore do not need to be considered in particular. 
Moreover, for poisson photon number distributed light fields, such as coherent light, dark counts during a measurement period do not alter the shape of the click statistics, it remains a binomial distribution.
Since the time interval of a measurement period is the total length of a detector output pulse, around $10\,\nano\second$, the dark count rate was about $10^{-3}$~counts, thus technically irrelevant.

Due to the dead time, dark counts occurring prior the light pulse can occupy pixels before the actual measurement starts, which makes these pixels unavailable for light detection. These preclicked pixels virtually reduce the overall available pixels $N$ of the PPD during the measurement. The impact to the click statistics may be insignificant for small and medium click numbers, but increasing the intensity high enough, almost every pixel will be triggerd. At this point occasionally reduced pixel numbers by preclicked pixels have a significant effect to the recorded click statistics. 
In that case, the statistics of preclicked pixels, which are depending on dark counts, stray light and dead time, affect statistical values more than the photon statistics of the light under investigation. We identified the preclicked pixels as the main reason why we were unable to record click statistics with the possible maximum click numbers of 100 clicks.
Furthermore, our investigations have shown that preclicked pixels start to have a significant influence to the moments of click statistics for photon numbers around 2.0 per pixel. 
Due to this we constrain our conclusions to photon numbers up to 1.5 per pixel.




Crosstalk describes the triggering of pixels which is not caused by the incoming light itself, but by the interaction of not triggered pixels with pixels triggered by photons. It is a known problem for assemblies of detectors with avalanche multiplication based detection~\cite{1503521,4447292}.
Crosstalk is correlated with the bias voltage and the spatial density of detectors in the system.
Our analysis reveal that, as implied in section: Nonclassicality indication, crosstalk has its greatest impact to the $Q_\mathrm{B}$-parameter at low intensities, where these devices are most commonly used.

Furthermore, the crosstalk parameter we introduced in the Methods section does not need to be fitted, but can be obtained by the extrapolation of $Q_\mathrm{B}$ for $|\alpha|\rightarrow0$. Since $Q_\mathrm B\stackrel{|\alpha|\rightarrow0}{\rightarrow}Q_\mathrm M$ and $c_m(|\alpha|\rightarrow0,\chi)=c_1(\alpha)\sum_{k=1}^mc^\mathrm{cross}_{m-1}(1,\chi)=c_1(\alpha)C_{m,\chi}$ we obtain
\begin{align}\notag
	\lim_{|\alpha|\rightarrow0}Q_\mathrm{M}&=\frac{c_1(\alpha)\sum_{k=0}^Nk^2C_{k,\chi}}{c_1(\alpha)\sum_{k=0}^NkC_{k,\chi}}\\
	&\hphantom{==}-\frac{{c_1(\alpha)}^2(\sum_{k=0}^NkC_{k,\chi})^2}{c_1(\alpha)\sum_{k=0}^NkC_{k,\chi}}-1\,,\\
	&=\frac{\sum_{k=0}^Nk^2C_{k,\chi}}{\sum_{k=0}^NkC_{k,\chi}}-1\,,
\end{align}
which is only depending on crosstalk.
Not only does this allow to determine of the crosstalk strength parameter without the input of any light click statistics, $Q_\mathrm B$ or $Q_\mathrm M$ itself can be used, for $|\alpha|\rightarrow0$, as a crosstalk quantity if the actual crosstalk distribution is unknown.~\cite{Du2008396}.

\section*{Discussion and Conclusion}
We demonstrated that the combination of click detectors in arrays (PPDs) and the processing of the output under consideration of the click counting distribution can have some major advantages compared to state-of-the-art photon counters.

The PPDs have a permissible intensity range which is nearly two orders of magnitude higher compared to their common usage, when the attempt of obtaining the photon statistics is given up for the benefit of click statistics. It has been shown, that the desired information about the quantum state can be obtained directly by observation of the click statistics, rather than the photon statistics.
Furthermore, even if future true photon number detectors would resolve as many photons as click detectors resolve clicks -- at present this is at least one order of magnitude higher -- click detectors have a much higher upper limit for the intensity of the signal field. The number of absorbed photons can easily exceed the number of resolvable detection events/overall pixels of the detector, while the crucial information is still preserved in the click statistics. This circumstance is valid not just for APD arrays but all click-detection based photon counters, such as time multiplexed setups, superconducting nanowires or frequency up-conversion.
We demonstrated measurements with more than 150 photons (mean value) with a resolution of almost all possible 100 clicks and we were able to deduce quantum state information of the light via the $Q_\mathrm{B}$-parameter.

Note that APD arrays with 400 APDs and 1600 APDs on one square millimeter are already commercially available~\cite{yamamoto2007,Haba2008154}.
It is easily possible to combine these detectors in arrays of PPDs (also commercially available by now) each with independent signal analysis, thus allowing to reach a five-digit click resolution. Although the APD fill factor of these devices, meaning how much of the detector area is actual photosensitive area, technically reduces the detection efficiency, in future this may be compensated by the use of microlens arrays.

Due to the widespread implementation and development of various click detector technologies, almost any demand on detector performances (efficiency, timing jitter, dark counts, etc.) can be met by commercially available solutions by now. At their current status, click detector systems are already a convenient alternative to most true photon counting detectors, which often require a sophisticated experimental setup, cryogenic cooling and professional expertise in both. The further increase of application potential due to the investigations on click detection amplify these advantages further~\cite{PhysRevA.95.033806}.

Our method is essential for states of which the magnitude of nonclassicality is very vulnerable to attenuation, e.g. squeezed states.
Furthermore, the compactness, which owns the PPDs a unique flexibility and versatility to changing experimental demands, and their improved intensity range make them a candidate for closing the gap between conventional single photon detection and linear detection with ordinary photodiodes.
Revealing quantum correlations and quantum state engineering with click detectors, explicitly based on the click counting distribution, have been done or proposed by now.~\cite{PhysRevLett.115.023601,sperling13,sperling14}
In future applications click detectors may adopt further tasks of true photon counting detectors, e.g. measuring phase space distributions~\cite{Eisenberg2004,WallentowitzVogel96}.
%

\section*{Acknowledgements}
This work was granted by the  Deutsche Forschungsgemeinschaft trough the SFB 652 (project B2 and in cooperation with projects B12 and B13).


\end{document}